\documentclass[conference]{IEEEtran}
\ifCLASSINFOpdf
  \usepackage[pdftex]{graphicx}
\else
  \usepackage[dvips]{graphicx}
\fi
\usepackage[nocompress]{cite}
\usepackage{amsmath,amssymb,amsfonts}
\usepackage{graphicx}
\usepackage{textcomp}
\usepackage{xcolor}
\usepackage{notoccite}
\usepackage[english]{babel}
\usepackage[utf8x]{inputenc}
\usepackage{amsmath,amsthm}
\usepackage{multirow, array}

\usepackage{fancyhdr}
\usepackage[normalem]{ulem}
\usepackage[hyphens]{url}
\usepackage[final]{microtype}
\usepackage{flushend}
\usepackage[bookmarks=false,breaklinks=true,letterpaper=true,colorlinks,linkcolor=black,citecolor=blue,urlcolor=black]{hyperref}

\usepackage{xcolor}
\usepackage{listings}
\usepackage{color}
\usepackage{xparse}
\usepackage{courier}
\usepackage{subcaption}
\usepackage{algorithm2e}

\usepackage{tabularx}

\newcommand{\ignore}[1]{}
\NewDocumentCommand{\codeword}{v}{%
\texttt{\textcolor{blue}{#1}}%
}

\definecolor{dkgreen}{rgb}{0,0.6,0}
\definecolor{gray}{rgb}{0.5,0.5,0.5}
\definecolor{mauve}{rgb}{0.58,0,0.82}

\lstdefinelanguage
   [x64]{Assembler}     
   [x86masm]{Assembler} 
   {morekeywords={real_fir, complex_fir, adaptive_fir, vector_dot, iir, %
   				  lbeg, lend, vector_add, vector_max, fft_256, dct_64, correlation,
                  r8,r8d,r8w,r8b,r9,r9d,r9w,r9b, %
                  r10,r10d,r10w,r10b,r11,r11d,r11w,r11b, %
                  r12,r12d,r12w,r12b,r13,r13d,r13w,r13b, %
                  r14,r14d,r14w,r14b,r15,r15d,r15w,r15b}} 

\fancypagestyle{firstpage}{
  \fancyhf{}
\setlength{\headheight}{10pt}

  \pagenumbering{arabic}
}

\title{HTS: A Hardware Task Scheduler for Heterogeneous Systems}

\author{Kartik Hegde$^{\dagger}$,\thanks{$^{\dagger}$Authors contributed equally to this work.}
        Abhishek Srivastaval$^{\dagger}$,
        Rohit Agrawal$^{\dagger}$\\
        University of Illinois at Urbana-Champaign\\
        \{kvhegde2, as29, rohita2\}@illinois.edu}
        
\IEEEoverridecommandlockouts
\begin{document}
\maketitle
\thispagestyle{firstpage}
\pagestyle{plain}

\begin{abstract}
As the Moore's scaling era comes to an end, application specific hardware accelerators appear as an attractive way to improve the performance and power efficiency of our computing systems. A massively heterogeneous system with a large number of hardware accelerators along with multiple general purpose CPUs is a promising direction, but pose several challenges in terms of the run-time scheduling of tasks on the accelerators and design granularity of accelerators. This paper addresses these challenges by developing an example heterogeneous system to enable multiple applications to share the available accelerators. We propose to design accelerators at a lower abstraction to enable applications to be broken down into tasks that can be mapped on several accelerators. We observe that several real-life workloads can be broken down into common primitives that are shared across many workloads. Finally, we propose and design a hardware task scheduler inspired by the hardware schedulers in out-of-order superscalar processors to efficiently utilize the accelerators in the system by scheduling tasks in out-of-order and even speculatively. We evaluate the proposed system on both real-life and synthetic benchmarks based on Digital Signal Processing~(DSP) applications.  Compared to executing the benchmark on a system with sequential scheduling, proposed scheduler achieves up to $12\times$ improvement in performance. 
\end{abstract}

\section{introduction}

We are at a challenging juncture in computer architecture research, where the imminent death of Moore's law threatens to slow down the performance growth we are used to, and rapidly growing fields such as deep learning demand more compute power than ever. Also, single thread performance improvement has saturated and Amdahl's law prevents us from any further exploitation of parallel computing for performance boost.
Furthermore, general purpose computing chips are significantly disadvantaged in terms of power and performance efficiency for hardware acceleration of different demanding tasks\cite{cpusarebad}. 

Therefore, the recent trend has been to use application specific hardware accelerators to boost the performance of computing systems. The application specific nature of these accelerators enables them to exploit the already known knowledge of execution of the targeted algorithms and data-access patterns to gain substantial improvements. To give a perspective, \cite{cpusarebad} claims over 500x higher energy efficiency over the general purpose CPUs in the task of video decoding.

\begin{figure}[!t]
  \begin{centering}
  \includegraphics[width=\columnwidth]{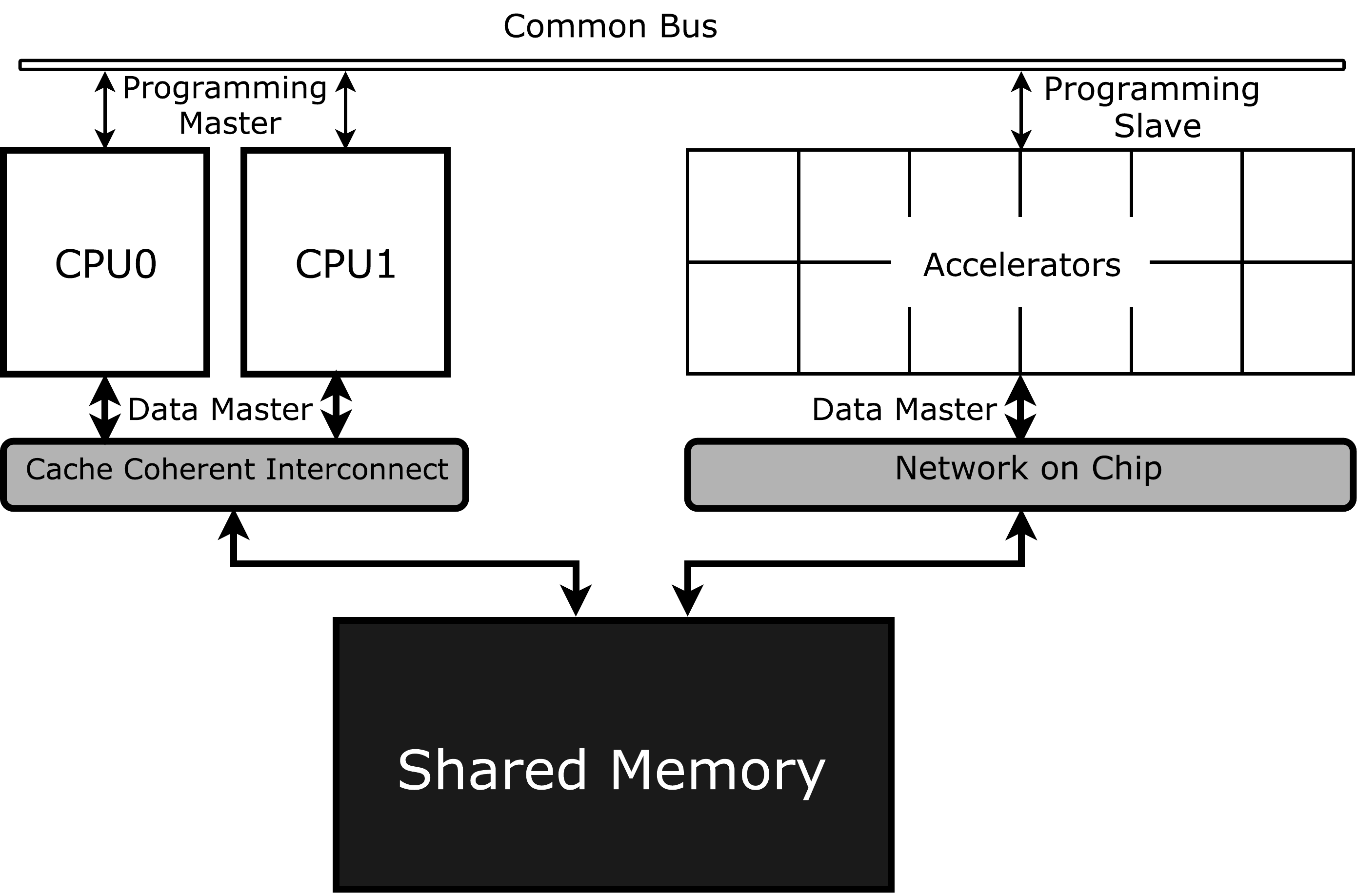}
  \caption{\label{fig:het_system}
            \footnotesize 
            A traditional view of a heterogeneous system} 
  \end{centering}
\end{figure}

However, it is non-trivial to build a massively heterogeneous system with a sea of accelerators, as the entire stack of OS, compilers and schedulers will have to be redesigned. It is also challenging to recognize the granularity of the accelerators to support a wide variety of algorithms. Since accelerators might share data that the general purpose CPU is generating/consuming, coherency and consistency of the memory also pose a challenge. There are also difficulties in establishing a uniform Virtual Address translation mechanism. Figure\ref{fig:het_system} depicts a heterogeneous system containing CPUs and accelerators sharing the memory.

Specifically, we recognize the difficulty of run-time task scheduling in systems with multiple hardware accelerators: where a \emph{task} is an abstraction level for a set of instructions that define a primitive, which tends to repeat across the program. User threads can be made up of multiple tasks and those tasks might be best performed by some accelerators in the system. The threads can have fine-grained parallelism leading to lots of dependencies among the tasks, coarse-grained parallelism that requires minimal interactions, or data parallelism that makes them completely independent. Furthermore, control flow changes may change the dependency structure completely during run-time. This clearly indicates the necessity of a run-time task scheduler for such heterogeneous systems, and a rich line of works \cite{houston2008portable, bauer2012legion, kale1993charm++, charles2005x10} based on run-time APIs address this challenge.

As noted in \cite{bauer2012legion, etsion2010task}, we can clearly see that the task scheduling on a CMP system is very analogous to instruction scheduling on different functional units in a general purpose CPU. We extend this notion to multiple domain specific accelerators and general purpose CPUs in a heterogeneous system.  Usage of custom schedulers in accelerators has been explored before\cite{carter2013runnemede}, but the notion of hardware based task scheduling for heterogeneous systems is unexplored. 

Arguably, one of the most profound developments in CPU micro-architecture that boosted the performance to higher levels is the design of out-of-order (OoO) pipelines to exploit \textit{instruction level parallelism}. Based on this inspiration, we argue that breaking down the threads into multiple tasks and exploiting \textit{Task Level Parallelism} effectively is the next growth area for high performance computer architecture. As we see the advent of massively heterogeneous systems, we believe that this work can contribute towards improving the utilization of hardware accelerators in the system.


In this light, we propose to design a hardware task scheduler that interfaces the CPU with all the accelerators in the system.   Again, drawing deom the analogy of an OoO pipeline, here the compiler/OS running on the CPU pushes tasks on to the scheduler, proposed scheduler is akin to the CPU frontend which considers the fine-grained accelerators as the \textit{Execution units}. The hardware scheduler is aware of the status of each of the slave execution units, giving a unique advantage to the hardware scheduler to assign and execute tasks in an \textit{Out-of-Order} fashion with a global view and even execute tasks \textit{speculatively}.

\textbf{Paper Outline.} The paper is organized as follows. Section~\ref{sec:background} gives an overview of the current literature and outlines the requirements for a heterogeneous task scheduler. Section~\ref{sec:motivation} details the inspiration behind the idea and describes the core insight of this paper. Section~\ref{sec:arch} describes the architecture of the system, accelerators and the hardware task scheduler. Finally, Section~\ref{sec:eval} shows the preliminary results of our work.

\section{background}
\label{sec:background}

Researchers are always looking at opportunities at various levels of the computing stack to improve application performance. This generally involves exposing parallelism at some level. The field has seen massive developments in this front through exploitation of Instruction Level Parallelism (ILP), Thread Level Parallelism (TLP) and Data Level Parallelism (DLP). All of them coincided with monumental developments in computer architecture. The shift towards out of order scheduling in processors led to massive growth in ability to exploit ILP while the prominence of multi-cores made exploiting TLP important. The addition of vector units, and consequent proposal of accelerators like GPGPUs, led to effective utilization of DLP. 

Following the aforementioned trend, we feel the highly well-documented surge of hardware specialization paves way for the usage of \textit{Tasks} as units of computation, with exploitation of Task Level Parallelism (TLP) becoming another important aspect of application programming. This should allow users to expose scope for concurrency at the application level, which when coupled with appropriate scheduling strategies can be mapped to threads, and hence instructions. Adding another level of abstraction should also enable users to pass intricate details about the application algorithm which might not seem intuitive to underlying runtime system and hardware, like fine-grained dependency tracking, eager memory management etc. Note that the notion of defining computation through tasks is not a novel concept. In fact, it has seen substantial exploration in the recent past. We summarize some of the prominent efforts below.

\subsection{Runtime-based Task Parallelism}
Most of the prominent task-based parallel computing environments consist of two components - (1) task-parallel API and (2) task runtime system ~\cite{Thoman2018}. The former defines the way an application developer describes parallelism, dependencies, data distribution options amongst other things, while the latter acts as the basis for implementing the APIs. The runtime defines the efficiency and ability of the environment. It determines the target architectures supported, task scheduling objectives, scheduling methodologies, support for fault tolerance etc. A large number of task-based programming environments have been developed over the past decades, with even established languages like C++ integrating tasks for shared memory parallelism.

Cilk~\cite{cilk} language allows task-based parallel programming with work stealing based scheduling. OpenMP ~\cite{openmp} integrate tasks into their programming interface, while task parallelism based libraries such as Intel TBB ~\cite{tbb} have emerged. Aforementioned environments were built for shared memory systems. The past few years have seen task-parallel models being built for heterogeneous hardware, like StarPU ~\cite{starpu},  and distributed systems like Chapel ~\cite{chapel}, X10 ~\cite{charles2005x10}, HPX ~\cite{hpx} and Charm++ ~\cite{kale1993charm++}. In distributed setting, tasks are combined with a Global Address Space (GAS) programming model to form a distributed execution of a task-parallel program ~\cite{Thoman2018}.

Legion~\cite{bauer2012legion} is a data-centric programming language with task based runtime. A program instance is defined in terms of its logical regions, which express locality and independence of program data, and tasks. The runtime system uses distributed, parallel scheduling algorithm and  a mapping interface to control movement of data and place tasks on devices based on locality. It is aimed towards heterogeneous systems. 

In OmpSs programming model ~\cite{ompss}, user specifies tasks with their data dependences. The runtime performs dynamic task dependence analysis, followed by dataflow scheduling and out-of-order execution. 

In such systems, as one tries to scale applications onto many processors, many more tasks are required to make full usage of available hardware resources. However, it has been observed that for fine-grained tasks, the overhead of software based task scheduling and management is too high to maintain scalable performance ~\cite{perf_analysis_of_picos}. This is generally attributed to task launch overheads, especially for fine-grained computations. Such studies have sparked interest in hardware based task dependence management systems. 

\begin{figure*}[!t]
  \begin{centering}
  \includegraphics[width=0.6\textwidth]{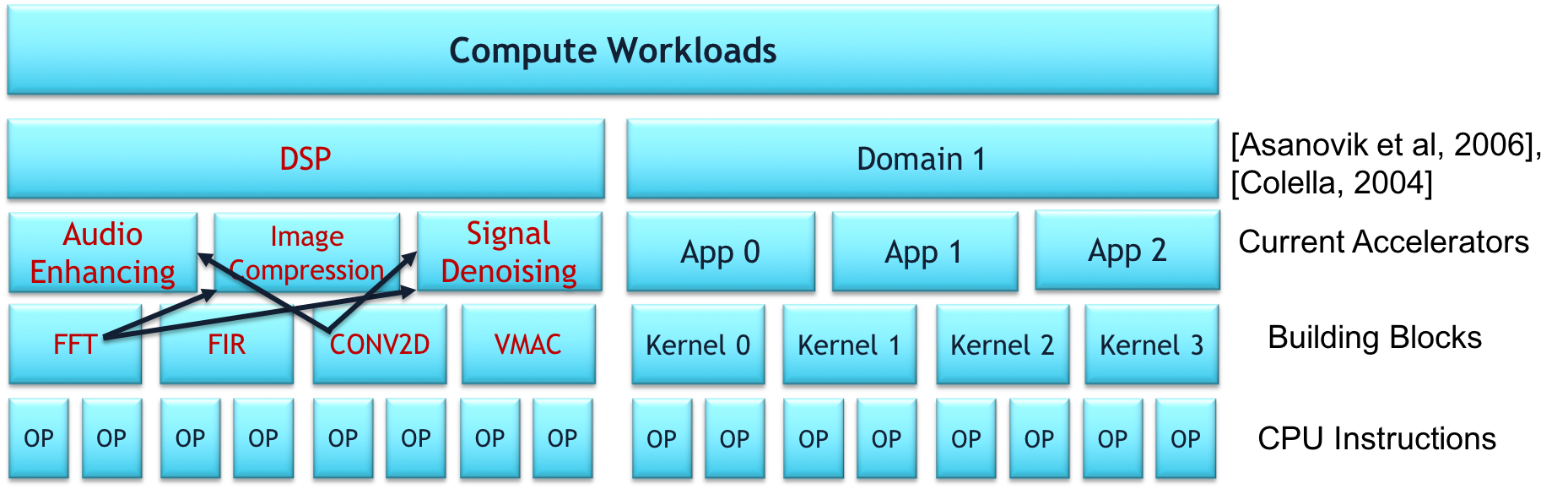}
  \caption{\label{fig:granularity}
            \footnotesize 
            Different abstractions levels for hardware acceleration} 
  \end{centering}
\end{figure*}

\subsection{Hardware-based Task Parallelism}
Task Superscalar ~\cite{etsion2010task} was proposed to accelerate task and dependence management using hardware. It showed promising results, but had issues with unresolved deadlocks due to queue saturation and memory capacity, which led to the proposal of a more enhanced design called Picos~\cite{picos}. It resolves these deadlocks and adds support for nested tasks. Carbon ~\cite{carbon} implements task queue operations and scheduling in hardware to support fast task dispatch and stealing. TriMedia-based multicore system ~\cite{trimedia} contains a hardware task scheduling unit, built on Carbon. TMU ~\cite{tmu} is a look-ahead task management unit for reducing task retrieval latency to accelerate task creation and synchronisation. Nexus++ ~\cite{nexus++} is a prominent contribution in this work. It, similar to Picos, leverages the work of dynamically scheduling tasks with real-time data dependence analysis but maintaining programmability of the system. 

We observe common denominations among all task-parallel programming environments. They require the programmer to present the application code in a new language, or at best annotate sections of the code for analysis. The former clearly affects portability, while latter might not give the runtime enough information to extract fine-grained parallelism. Also, to the best of our knowledge, none of aforementioned task-parallel models with software/hardware scheduler caters to heterogeneous systems with specialized accelerators. Our survey findings suggest decoupling programming and task management aspects of a task-parallel programming model to enable portability, and to develop task management hardware for modern day systems. 

Runnemede~\cite{runnemede} is a notable contribution in the literature,  which echoes many of the design principles of our proposal. It is a co-designed hardware/software effort where hardware, execution model, OS/runtime and applications are being simultaneously developed. Its execution model divides programs into tasks called \textit{codelets}~\cite{runnemede}, which are self-contained units of computations with defined inputs and outputs. Notably, it provides a number of programming models with different tradeoffs, which is one way of enabling portability. The user can use higher-level models like Hierarchically-Tiled Arrays~\cite{hta} and Concurrent Collections~\cite{cnc}, or code to runtime's codelet model to get a lower-level interface to the hardware. 

The execution model is based on a dataflow model. Carter et.al~\cite{runnemede} elucidate the following characteristics of a dataflow execution model, which make it well suited to extreme-scale systems, similar to what we intend to work on:
\begin{enumerate}
\item{It allows easy exploitation of all parallelism within each phase of an application, not requiring its thread based division}
\item{It incurs less synchronization costs as only producer and consumer(s) of an item need to synchronize}
\item{It enables clear break down and efficient scheduling of tasks onto different parts of the system in a non-blocking manner, which makes it easy to schedule code close to its data, marshal input data at the location where the computation is to be performed etc}
\end{enumerate}

Runnemede architecture designs two types of cores - general purpose Control Engines (CE) and specialized Execution Engines (XE) intended for execution of codelets. Note that an XE cannot perform I/O operations, instead an I/O operation is represented as a dataflow dependence between codelets which is performed by a CE.

\section{Motivation}
\label{sec:motivation}

It is well established that the application specific hardware accelerators significantly outperform CPUs and GPUs in a variety of tasks such as Deep Learning, genome sequencing, computer vision, digital signal processing etc,. These accelerators are generally coupled with an API and upon the CPU's command, they complete the task they are built for. Unfortunately, these hardware blocks can not be used for anything other than the application that it is built for. 


However, there can be a mid-point between a fully general purpose system and an inflexible hardware accelerator. \cite{dwarfs} proposes several dwarfs in computing algorithms that form the basic categories of computations that generally occur. From an overview, it seems like building a system that does well on each of these dwarfs should provide performance uplift. But, that is simply not the case, largely because workloads can not be segregated into to dwarfs at that abstraction, but rather requires much lower abstraction. For e.g., image processing applications fall into dense and sparse linear algebra, but it is too high level of an abstraction to meaningfully accelerate them in hardware.

We provide a core insight here: \emph{a massively heterogeneous system should have a large number of accelerators at a lower abstraction such that they are usable as a basic functional block across a large range of applications simultaneously.} Figure~\ref{fig:granularity} depicts this pictorially. Every application is made up of several kernels, and each kernel would need several functions to be implemented and at the lowest level of the hierarchy are OPs(operations) that constitute a function. Figure~\ref{fig:granularity} gives an example of image processing application for better clarity. While most hardware accelerators today are at Application granularity (for example, Deep Learning inference) and CPUs at basic OPs granularity, we argue that building a large number of accelerators at the kernel granularity enables them to be reused across s across several applications of a domain.

To enable a large number of accelerators developed to execute various functions to be shared across a number of applications and kernels, we require a run-time scheduling system. \textit{Interestingly, such a scheduler is very similar to a hardware scheduler in an out-of-order processor, because each of these accelerators execute a task associated with a memory region, akin to instructions with operands}\cite{bauer2012legion}. The complexity of such a scheduler is non-trivial, because it has to deal with many of the difficulties that the OoO processor scheduler has to deal but at a much higher granularity. 

Several earlier works \cite{bauer2012legion,kale1993charm++} proposed software approaches to scheduling on CMPs, and a reasonable extension of these works should enable a massively heterogeneous system task scheduler. In such cases, as depicted in Figure~\ref{fig:het_system}, CPUs interact with the accelerators as masters on a common programming bus. Accelerators respond to requests made by the masters via an interrupt to the requester. The accelerators would be masters on the databus to access the memory like CPUs. However, such software task scheduling has inherent disadvantages: 
\begin{enumerate}
	\item The instance of scheduler running on the host CPU is an overhead \cite{bauer2012legion}
    \item The task completion signals will have to be conveyed via interrupts which can have a long latency.
    \item A central manager of the hardware accelerator is absent, which makes accelerator management requirements such as Dynamic Voltage-Frequency Scaling (DVFS) difficult.
\end{enumerate}

\begin{figure}[!t]
  \begin{centering}
  \includegraphics[width=\columnwidth]{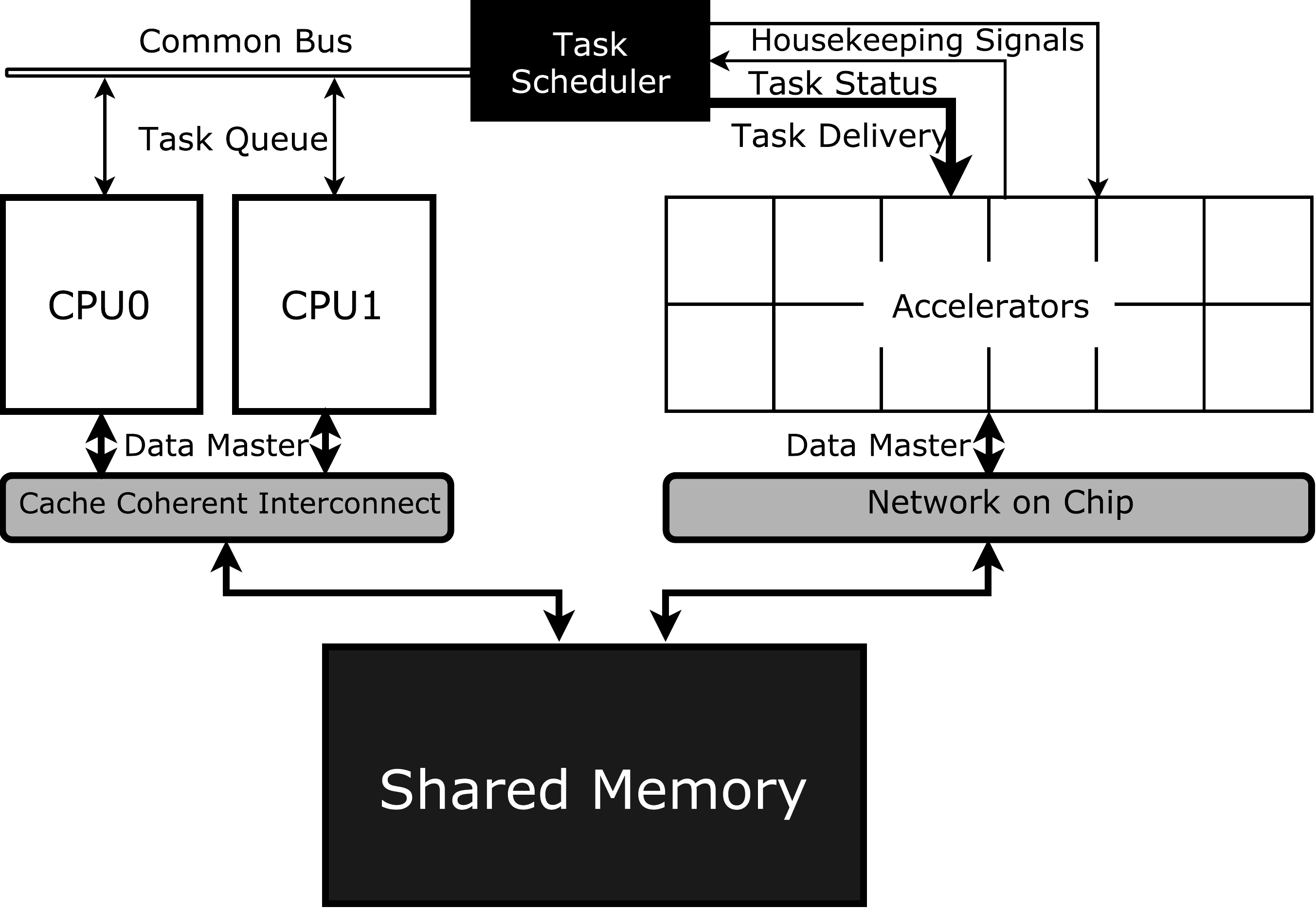}
  \caption{\label{fig:system}
            \footnotesize 
            Proposed Task Scheduler integrated in the system} 
  \end{centering}
\end{figure}

However, all the above mentioned disadvantages can be alleviated if the system contains a central hardware task scheduler. Figure~\ref{fig:system} modifies Figure~\ref{fig:het_system} to include a Hardware Task Scheduler (HTS) in the system that interfaces the CPUs and the accelerators. 
CPUs can push new tasks and the associated meta-data to the HTS and continue execution or poll on the scheduler in case of a dependency. HTS maintains a queue of tasks and the associated metadata, akin to the instructions to be executed in an OoO CPU core. 
HTS is aware of the busy status of each of the slave accelerator in the system, based on which it can schedule the tasks. Accelerators notify the HTS once the assigned task is complete via a physical signal, which is orders of magnitudes faster than interrupts. 
HTS can also control the power-management of all the accelerators in the system.

Most importantly, HTS can schedule tasks in an out-of-order fashion based on dependencies and status of the accelerator, which can bring a great speed-up in a task-parallel system. 
Furthermore, speculatively executing tasks based on both control and data branches can further bring a significant boost in performance.
Hardware instruction schedulers have revolutionized the modern micro-processor design and has greatly simplified the OS and compilers associated.
The key take-away point here is that similar enhancements in OoO and speculative task execution mechanisms based on hardware are necessary to realize a massively heterogeneous system architecture.

\section{Architecture}
\label{sec:arch}

In this section, we detail the overall architecture of the proposed heterogeneous system and the task scheduler. We will first describe the overall system design and then elucidate the accelerator and scheduler interfaces. 

\subsection{System Design}
\label{subsec:sys_design}

There are several crucial design decisions to make to design a system with multiple accelerators. 
\begin{enumerate}
\item Are accelerators masters on the data-bus? Do they have an internal DMA controller?
\item Are accelerator scratch-pads coherent with the system?
\item How can CPUs program the accelerators?
\item How are interrupts configured?
\item How to manage contention?
\end{enumerate}

We largely agree with the system design proposed in \cite{parade}, where each accelerator has its own DMA engine. The scratchpads are not coherent with the system memory and require explicit synchronization via barriers. A centralized manager manages the tasks to be scheduled on the accelerators~(referred to as GAM in \cite{parade}), and handles notifying the CPUs when the task is complete. Managing contention is currently beyond the scope of this work and we leave it for future work. An overview of the proposed system is shown in Figure~\ref{fig:system}

\begin{figure}[!t]
  \begin{centering}
  \includegraphics[width=0.65\columnwidth]{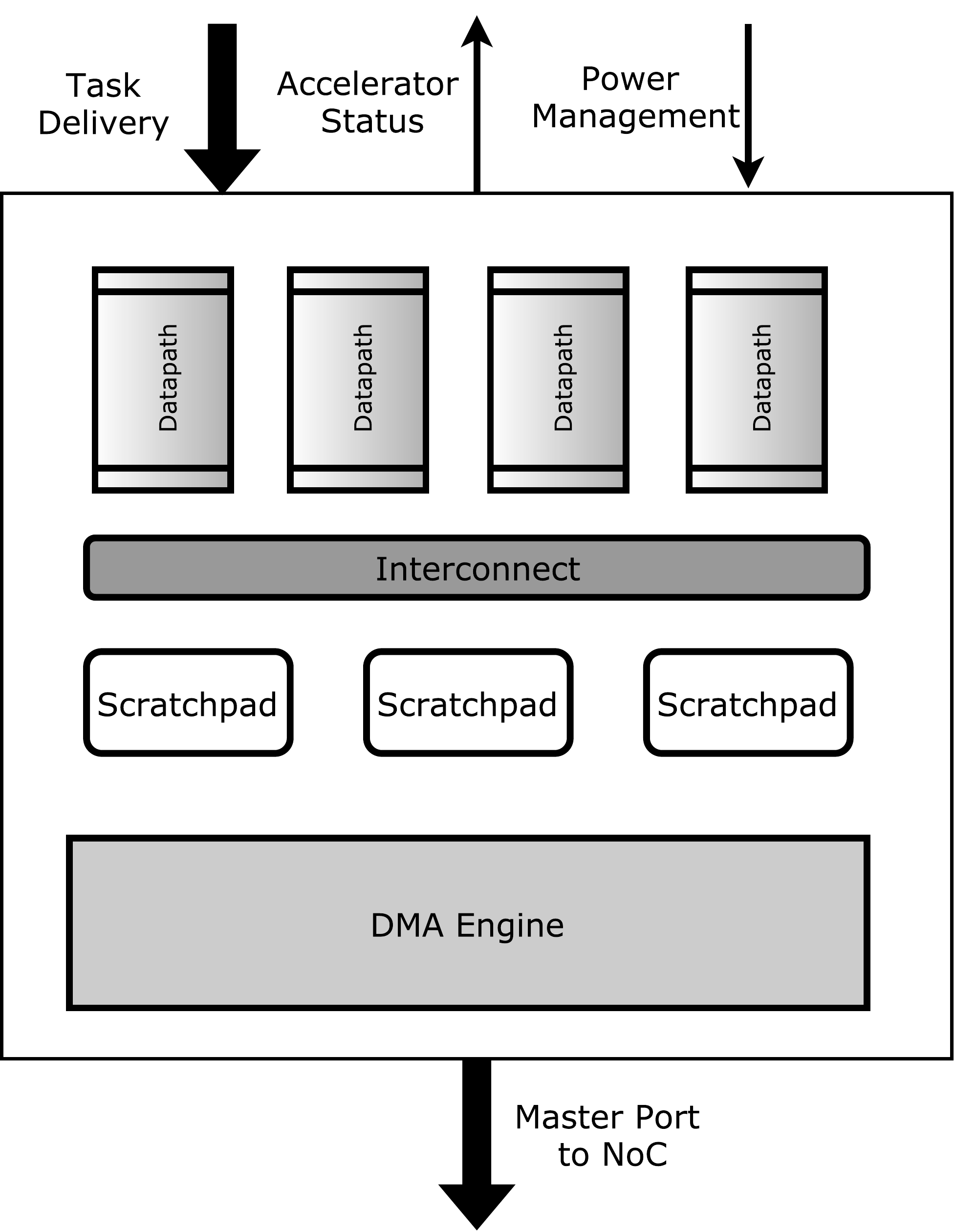}
  \caption{\label{fig:acc}
            \footnotesize 
            A basic interface for a candidate accelerator in the system} 
  \end{centering}
\end{figure}

\subsection{Accelerator Design}
\label{subsec:acc_design}
When the system contains a large number of accelerators, it is essential to maintain a homogeneous external interface for each accelerator. Figure~\ref{fig:acc} depicts the interface of every accelerator in the system. Each accelerator contains a DMA engine that can fetch the data from the memory via the Master port connected to the NoC. The scheduler can deliver the data to the accelerator via Task Delivery slave port, which should contain the entire description on the task to be performed and the base address. Accelerator can convey the \emph{busy} status to the manager via \emph{accelerator status} signal. The power management of the accelerator is again controlled by the manager.

\begin{figure*}[!t]
  \begin{centering}
    \includegraphics[width=0.75\textwidth]{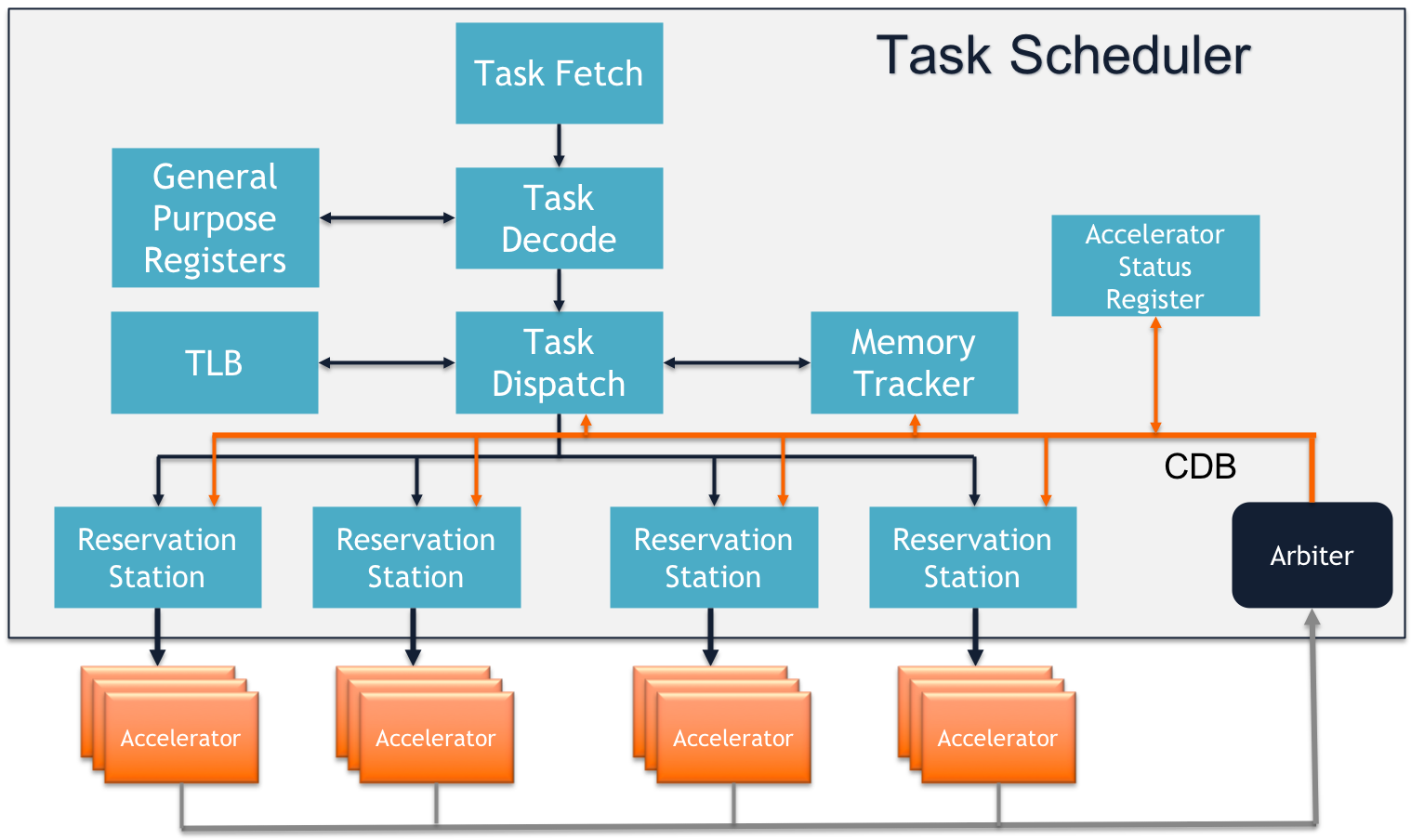}
  \caption{\label{fig:scheduler}
            \footnotesize 
            Proposed Task scheduler that retains the basic design of an OoO processor} 
  \end{centering}
\end{figure*}

\subsection{Task Scheduler}
\label{subsec:sched_design}

The core idea of having a large number of accelerators that are shared across applications is realizable only when the run-time scheduling is realized in the system to effectively share the resources. 
As proposed in Section~\ref{sec:motivation}, we design a \emph{Hardware Task Scheduler}~(HTS) that receives tasks from CPUs and manages the task-run on the accelerators. We develop the task scheduler as an OoO core that is capable of scheduling tasks on the accelerator dynamically. The key insight here is that all the optimizations that CPU development has seen can be brought to task scheduling with minimal changes.

Figure~\ref{fig:scheduler} gives a top level view of the proposed HTS. As can be observed, its design is completely based on an OoO processor that can execute instructions in an out-of-order fashion and is able to execute instructions speculatively. We exploit the fact that tasks can be executed in an OoO fashion as well\cite{bauer2012legion}, where the user's program is compiled to generate a task-flow graph~(explained in Section~\ref{sec:prog_model}). Each CPU running the program, therefore, can push the tasks to the scheduler and be notified of completion of the tasks via a dedicated bus.

\subsubsection{Overview}
The HTS receives tasks from the CPUs into the \emph{Task Queue}, which is then decoded by the \emph{Task Decode} logic. The decoded tasks contain information on the type of the task and the associated meta-data~(Section~\ref{sec:prog_model}). Every task is associated with a memory region and there are likely to be dependencies among them. This is analyzed by the \emph{Task Dispatch} logic, which can re-order the tasks in a window that is decided at the design time~(similar to instruction window). These tasks are dispatched according to an OoO issue logic, which are named as \emph{Reservation Stations} since its functionality is similar to Reservation Stations as proposed by Tomasulo's algorithms for OoO scheduling~\cite{tomasulo}. Based on whether the accelerator that the task maps to is free~(similar to busy status of the Functional Units in a CPU), \emph{Reservation Station} dispatches the tasks to the accelerators. Note that the width of dispatch is a design parameter.

The accelerators receive these tasks and proceed to perform them. Note that accelerators are masters on the data-bus, hence they can fetch the data required for the task. Once an accelerator completes the task and writes the result back to the memory, it sends the completion status back the HTS. This is conveyed to the HTS via a \emph{Common Data Bus}~(CDB), which clears other dependencies waiting for this task to complete. Every cycle, the reservation stations can issue tasks whose dependencies are cleared based on the availability of the accelerator. This enables the design to issue multiple tasks per cycle, hence similar to a \emph{Superscalar} design.

\subsubsection{Resolving RAW dependency}
One of the most common type of hazard occurring in an OoO pipeline is a Read-After-Write~(RAW) hazard. Our design resolves this in the Task Dispatch stage using an additional structure named \emph{Memory Tracker}. Each incoming task is assigned an ID by the dispatch. The dispatch logic informs the Memory Tracker about the output memory region that the outgoing task is going to write, and the corresponding task ID is stored. When a new task is ready for dispatch, the input memory region of the new task is scanned in Memory Tracker for any dependencies. In case an entry is found in the Tracker, corresponding task ID is returned and the new task is dispatched to the reservation where it waits till the dependency is resolved. The completion of each task has to announced on the CDB which is controlled using an arbiter that implements a ticket lock system for serialization. 

\subsubsection{Speculative Execution}
\label{subsubsec:speculation}
Interestingly, the whole concept of \emph{Speculative Execution} can be applied in task scheduling. The dataflow graph is executed speculatively whenever a branch is encountered. The main challenge in realizing speculation in task scheduling is that the results of the speculative tasks can not be reversed since they operate on the memory directly. We get around this by allocating a region of memory for speculated tasks to operate on, which can be discarded in case of mis-speculation. This is in similar spirit of how \emph{Transactional Memory} operates.

We consider three different types of branches to speculate upon:
\begin{enumerate}
\item \textbf{Register-Read~(RR):} These branches can be resolved by simply accessing the general purpose register bank in the scheduler. This causes a single cycle bubble, and comparing to the cycles that each task takes~(typically in 1000s of cycles), it is not beneficial to speculate. We simply incur the bubble cost and resolve the branch.
\item \textbf{Memory-Read~(MR)}: These branches are based on the data in some location in memory. This requires spawning a new task to read memory which can potentially take a large number of cycles. Hence, we consider these for speculation.
\item \textbf{Bus-Read~(BR)}: These branches depend on the output of some task that is yet to finish. In this case, the dispatch unit continues the execution speculatively but continues to monitor the CDB to resolve the branch.
\end{enumerate}

When the HTS is running in speculative mode, each task's output is mapped to a new location in the \emph{Transactional Memory}, which is a dedicated part of memory reserved for speculative tasks. The dispatch unit queries the \emph{Task Lookup Buffer}~(TLB) to allocate a new region for the output of the current speculative task, and the corresponding mapping in stored in the TLB. Additionally, each subsequent task's input memory region is looked up in the TLB to get the mapping, if present. Note that, each speculation is given an ID and the same is noted for each mapping present in the TLB. Based on the branch resolution, there can be two cases:
\begin{enumerate}
\item \textbf{Mis-speculation}: In this case, all the entries corresponding to the speculation ID is discarded from the TLB. All the tasks that are evicted from the TLB are aborted immediately. Also, we need not operate on the Transactional Memory, because the deletion of entries in the TLB is equivalent to invalidating/erasing the mis-speculated region of the TM.
\item \textbf{Correct Speculation}: In this case, execution can continue normally and the mapping is retained in the TLB. Any future tasks that request the data present in the region that was speculated and is present in the TLB, is remapped to read the data from the TM. Note that, if the TLB/TM become full, the HTS will stall and delete the mapping by copying the data from TM to the mapped location in memory. This ensures functional correctness of the HTS.
\end{enumerate}

\subsubsection{Accelerator Status}
One important requirement of the HTS is to be able to monitor the busy status of each accelerator in the system. We achieve this by maintaining a directory named \emph{Accelerator Status Register}~(ASR) as depicted in Figure~\ref{fig:scheduler}. The reservation station checks the ASR before releasing any task to the accelerator to make sure the required accelerator is idle. The ASR also monitors the CDB to clear the busy status of any accelerator that completed the task.

\subsubsection{General Purpose Registers}
HTS provides a \emph{General Purpose Register}~(GPR) bank for supporting different programming models. We elaborate the programming model in Section~\ref{sec:prog_model}. The number of registers in the GPR is a design time parameter. Each register can be addressed as $Rx$, where $x$ is the number of the register.

\section{Programming Model}
\label{sec:prog_model}

\begin{figure}[!t]
  \begin{centering}
  \includegraphics[width=0.6\columnwidth]{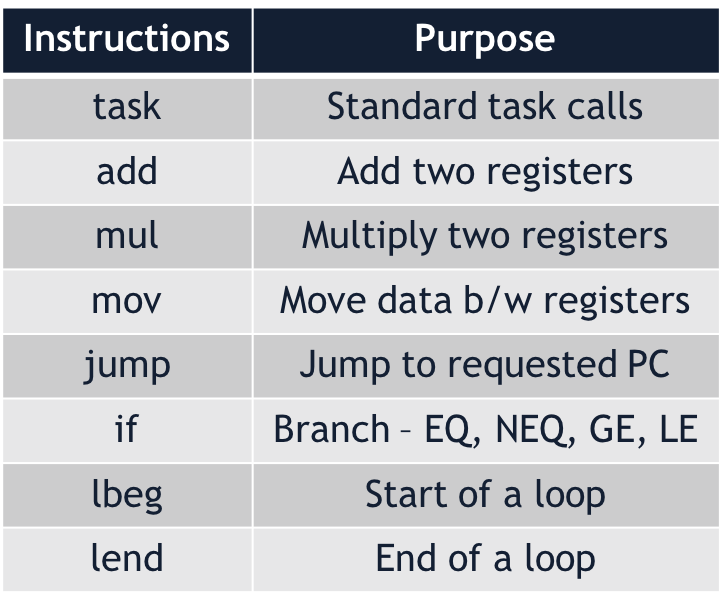}
  \caption{\label{fig:isa}
            \footnotesize 
            Supported instruction set architecture for programming the dataflow graph} 
  \end{centering}
\end{figure}

As described in Section~\ref{sec:arch}, the HTS executes dataflow graph. However, the CPU or the compiler need to describe the dataflow graph that can be understood by the HTS. Again drawing from the analogy of CPUs and high level languages to describe program, we need a unifying set of rules and instruction with which all the dataflow graphs can be described. We need a mechanism to divide the application into tasks, accompanied by their data and control dependencies. Note that portability across such heterogeneous systems is an important design principle of our proposal. So, we wish to employ a generic programming model which describes the relationships among the different entities (tasks, data, control) and leave the job of scheduling and execution on the \emph{Task Scheduler} and heterogeneous system respectively. In this section, we describe our \emph{Instruction Set Architecture} and provide a glimpse into programming the HTS to execute dataflow graphs.
 
\subsection{Instruction Set Architecture}
\label{subsec:isa}
Figure~\ref{fig:isa} lists the instructions supported by the HTS. Along with the \codeword{task} instruction, that can be used to assign a task to an accelerator, we support arithmetic instructions like \codeword{add}, \codeword{mul} and \codeword{mov} that can be used to operate on the GPRs. In order to support all types of dataflow graphs, we also add \codeword{if} instruction for branches, \codeword{jump} to jump to a specific part of the dataflow graph, \codeword{lbeg} to start a loop and \codeword{lend} to end a loop. All the instructions are of 128 bit width and the breakdown of each field is given in Table~\ref{tab:isa}.

\begin{table}[h!]
\centering
\caption{Instruction Breakdown}
\label{tab:isa}
\begin{tabular}{||c| c||} 
 \hline
 Range & Purpose  \\ [0.5ex] 
 \hline\hline
 \textbf{[7:0]} &  Accelerator ID\\
 \hline
 \textbf{[23:8]} &  Input Memory Region\\
 \hline
  \textbf{[31:24]} &  Input Memory Size\\
 \hline
  \textbf{[47:32]} &  Output Memory Region\\
 \hline
  \textbf{[55:48]} &  Output Memory Size\\
 \hline
  \textbf{[59:56]} &  Task ID\\
 \hline
  \textbf{[63:60]} &  Process ID\\
 \hline
  \textbf{[67:64]} &  Control\\
 \hline 
 \textbf{[127:68]} &  Metadata~(for the accelerator)\\
 \hline
\end{tabular}
\end{table}

\subsection{Describing the dataflow graphs}
\label{subsec:program}
The HTS executes the program written in assembly language. Each accelerator is given a keyname~(for e.g., \codeword{fft_256} for all the accelerators that can execute a 256 point FFT). The keyname would be assigned an accelerator ID when the code is compiled. After the keyname, the instruction should be described as mentioned in Table~\ref{tab:isa}. Each field is a hexadecimal number, and a simple dataflow graph depicting a set of independent nodes is described below.

\begin{lstlisting}
real_fir 10 2 13 2 0 0 0 0000
complex_fir 16 2 19 2 1 0 0 0000
adaptive_fir 23 3 28 3 2 0 0 0000
vector_dot 40 4 48 4 3 0 0 0000
iir 32 3 36 3 4 0 0 0000
\end{lstlisting}

\subsection{Support for Loops}
\label{subsec:loop}
The dataflow graphs need to support looping, as it is widely found in real-life applications. HTS recognizes the start of a loop with \codeword{lbeg} instruction and begins a counter based on the requested number of iterations. The \codeword{lend} instruction depicts the end of the loop body and the associated loop count register. HTS loops through the loop body for requested number of iterations as shown in the example program below.

\begin{lstlisting}
mov 58 0 2 0 1 0 0 0001
mov 3 0 3 0 2 0 0 0001
mov 75 0 6 0 3 0 0 0001
lbeg 4 4 0 0 4 0 0 0001
add 4 2 5 0 5 0 0 0001
add 4 6 7 0 6 0 0 0001
iir 5 3 7 3 7 0 1 0000
lend 0 4 2 0 8 0 0 0001
\end{lstlisting}

\subsection{Support for Branches}
\label{subsec:branch}
One of the most important part of the ISA is its support for branches, based on which the speculative execution is supported. The \codeword{if} instruction depicts the start of a branch and it describes the dependency of the branch that helps the HTS classify the type of branch~(Section~\ref{subsubsec:speculation}). The \codeword{if} instruction should also program the PC jump that the HTS has to perform if the branch is taken. Below is an example code supporting a branch, where the branch condition is evaluated based on the memory region $93$~(hence an MR branch) which requests a PC jump by $18$ if the branch is taken. Note that the branches can co-exist with loops as well.

\begin{lstlisting}
mov 3 0 a 0 0 0 0 0001
real_fir 10 2 13 2 0 0 0 0000
complex_fir 16 2 19 2 1 0 0 0000
if 93 a 12 0 1 0 d 0000
adaptive_fir 23 3 28 3 2 0 0 0000
iir 32 3 36 3 3 0 0 0000
vector_dot 40 4 48 4 4 0 0 0000
vector_add 55 4 62 4 5 0 0 0000
vector_max 68 5 76 5 6 0 0 0000
fft_256 84 6 93 6 7 0 0 0000
dct_64 102 2 106 2 8 0 0 0000
correlation 110 3 115 3 9 0 0 0000
\end{lstlisting}

\section{Evaluation}
\label{sec:eval}
In this section, we describe our evaluation strategy and the current results in our simulation environment.

\begin{table*}[!h]
\centering
\caption{DSP Functions modeled as accelerators}
\label{tab:kernels}
\begin{tabular}{||c|c|c|c|c||} 
 \hline
 Kernel & Description & Input dataframe size & Cycles \\ [0.5ex] 
 \hline\hline
 \textbf{Real FIR} & Real input valued finite-duration impulse response filter	&  40 & 921\\
 \hline
 \textbf{Complex FIR} &	Complex input valued finite-duration impulse response filter &  40 & 3696\\ [1ex] 
 \hline
 \textbf{Adaptive FIR} & Least mean square finite-duration impulse response filter& 40  & 4384\\ [1ex] 
 \hline
   \textbf{IIR} &	Infinite-duration impulse response filter&  40 & 2450\\ [1ex] 
 \hline
  \textbf{Vector Dot} &	Calculates vector product of two vectors &  40 & 53\\ [1ex] 
 \hline
 \textbf{Vector Add} & Adds two vectors	& 40  & 131\\ [1ex] 
 \hline
 \textbf{Vector Max} & Computes largest value in a vector	& 40  & 55\\ [1ex] 
 \hline
 \textbf{FFT} &	256-point radix-2 in-place complex Fast Fourier Transform &  256 & 18673\\ [1ex] 
 \hline
 \textbf{DCT} &	Discrete cosine transform &  64 & 874\\ [1ex] 
 \hline
\textbf{Correlation} & Computes measure of similarity  & 40  & 753\\ [1ex] 
 \hline

\end{tabular}
\end{table*}
\subsection{Workload Characterization}
\label{sec:workload}

\begin{figure*}[!h]
  \begin{centering}
  \includegraphics[width=0.75\textwidth]{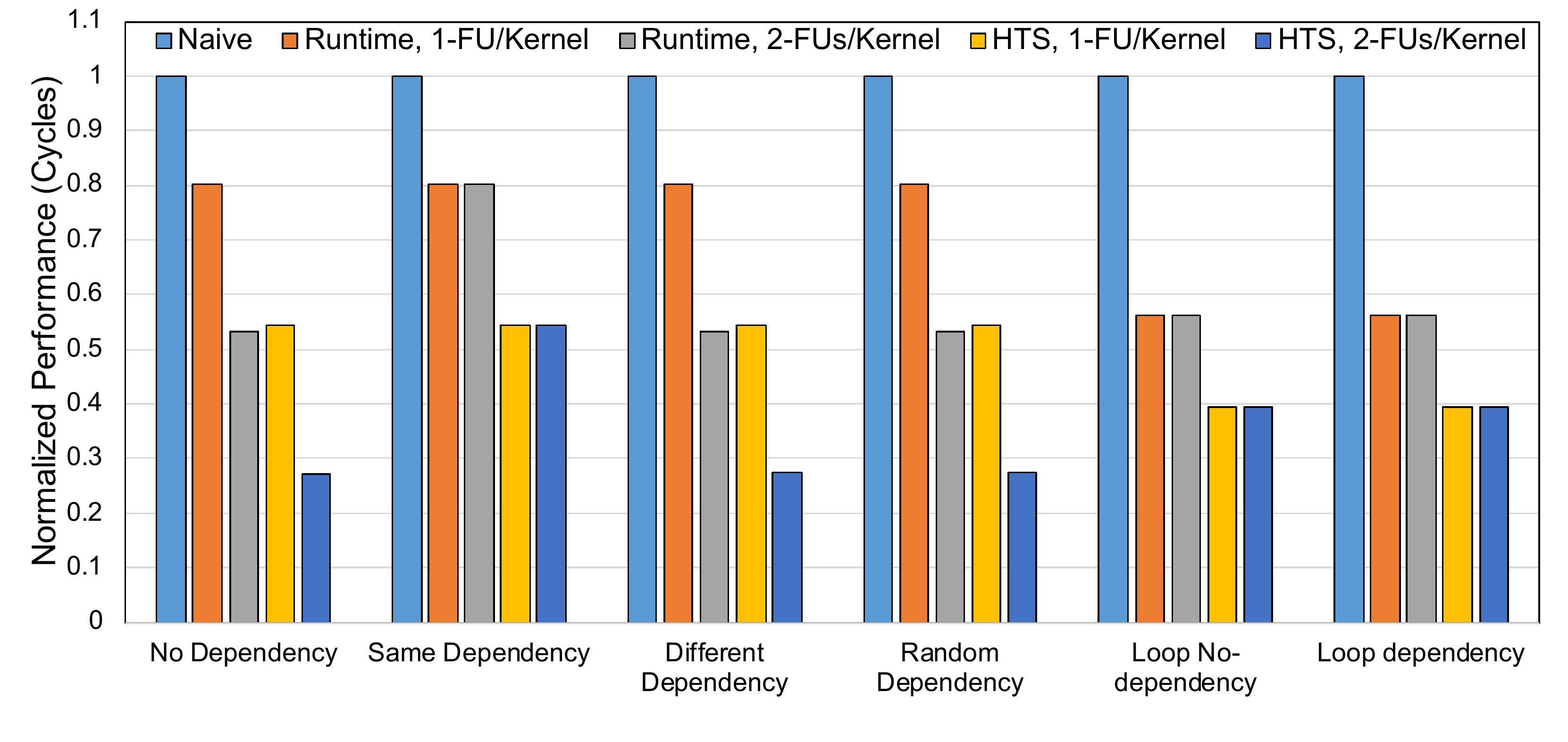}
  \caption{
             Performance comparison on synthetic benchmarks without branches.
             }
     \label{fig:custom1}
  \end{centering}
\end{figure*}

Generally, hardware accelerators are built for specific applications. A modern mobile SoC would contain accelerators for graphics processing, video decoding, digital signal processing etc. However, a massively heterogeneous system as described in Figure~\ref{fig:granularity} would contain a large number of Function level accelerators. Therefore, to demonstrate the effectiveness of such a system, we choose a workload based on several reasons:
\begin{enumerate}
\item There should be a large number of applications requiring hardware acceleration.
\item The applications should be decomposable into Kernels and Functions as described in Figure~\ref{fig:granularity}.
\item The applications should share the Kernels and Functions across each other.
\end{enumerate}

In this work, we demonstrate the advantages of building a massively heterogeneous system with a task scheduler based on Digital Signal Processing~(DSP) workloads. The choice is driven by the reasons described above. In addition, DSP workloads contain popular real-life applications, which are widely benchmarked by several previous works. It is also interesting to compare them with DSP processors, which form a mid-point between general-purpose CPUs and ASICs. 

\subsection{Accelerators}
\label{subsec:eval_acc_design}
We assume presence of accelerators for the Functions described in Table~\ref{tab:kernels}. These accelerators were benchmarked by Lennartsson et.al~\cite{dspaccelerators}. We use the mentioned benchmarking cycle numbers in our experiments. These cycle numbers are crucial for our further evaluation of the task scheduling systems.

Table~\ref{tab:kernels} provides an enumeration of DSP Functions that we model as accelerators in our system. For the sake of our experiments, we assume that the aforementioned Functions can be used to run any DSP Kernel/Application.

\subsection{Experiments}
We modeled the proposed Hardware Task Scheduler (HTS) in python. The implementation is cycle accurate. It assumes an incoming stream of tasks sent by the CPU. This is accomplished by passing an assembly (.asm) file to the model, which contains tasks described as per our ISA. The model is configurable by number of accelerators per Function.

Our experimentation can be divided into two sections - custom benchmarks and real application. We devised different custom-made benchmarks to observe the behavior of various features of our proposed HTS. We then pick audio compression, a real life application, to show the feasibility of our proposal and its observed performance. 

For every experiment, we compare three scheduling algorithms : 
\begin{enumerate}
\item{\textit{Naive scheduling} - The CPU schedules one task at a time (in-order). For each task, CPU schedules the task, and waits for its completion before processing to the next task. We estimate its performance by adding (execution cycle number, interrupt latency) for each task. Note that interrupt latency is independent of the task.}
\item {\textit{Runtime (Software) based scheduling} - An out-of-order Runtime running on the CPU schedules tasks. We design it as the manifestation of our HTS design in software. We estimate its performance by adding (software scheduling overhead, interrupt latency) for each task. We model software scheduling overhead as memory access latency if our exact HTS was implemented in hardware, in which case, Memory tracker, Reservation Station etc would actually reside in memory. We assume L2 cache hit for each memory access.}
\item {\textit{HTS Scheduling} - Our out-of-order, speculative hardware scheduler}
\end{enumerate}

We use ARM Cortex-A interrupt latency~\cite{interruptlatency} and ARM Cortex-A9 L2 cache memory access latency~\cite{memoryaccess} for our experiments. Also, performance is modeled by clock cycle numbers. Note that we are making crude estimations, especially for Runtime based scheduling, for the purpose of comparison. So, these experiment values are not absolute in any sense. Our intention is to provide a ballpark performance of other scheduling algorithms to shed light on the advantages of our proposed HTS. 

\begin{figure}[!t]
  \begin{centering}
  \includegraphics[width=\columnwidth]{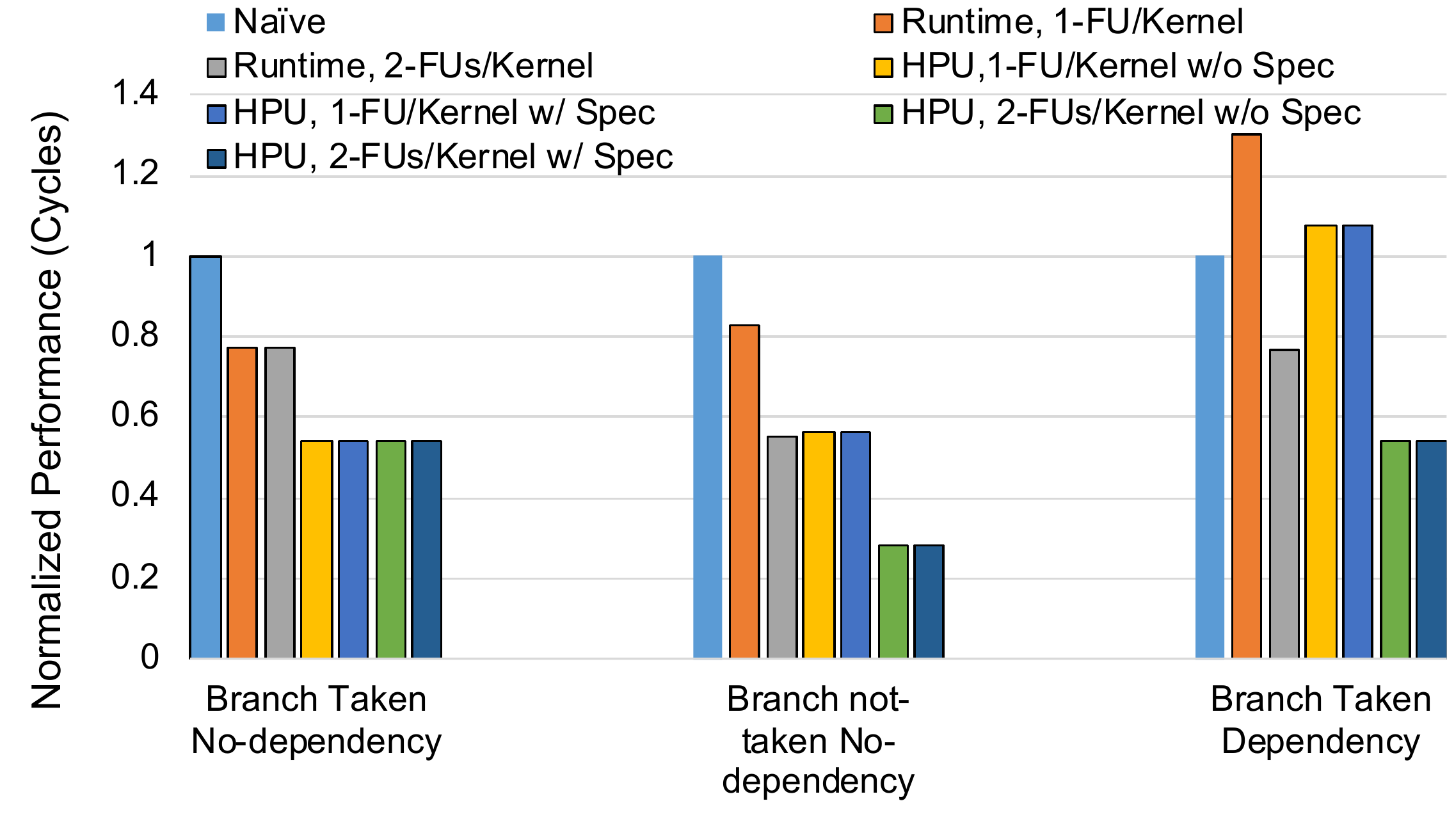}
  \caption{\label{fig:power_consumption}
            \footnotesize 
             Performance comparison on synthetic benchmarks with branches.
             }
    \label{fig:custom2}
  \end{centering}
\end{figure}

\subsubsection*{Custom-made Benchmarks}
We use the following custom-made benchmarks:
\begin{enumerate}
\item{No Dependency - no dependencies among any tasks, has no loops or branches}
\item{Same Dependency - dependencies only among tasks mapped to same functional unit (Ex - instance of FFT and FFT ), has no loops or branches}
\item{Different Dependency - dependencies only among tasks mapped to different functional units (Ex - instance of FFT and Correlation), has no loops or branches}
\item{Random Dependency - no definite pattern among dependencies, has no loops or branches}
\item{Loop No Dependency - one loop with no dependency outside the loop, has no branches}
\item{Loop Dependency - one loop with dependency of the loop iteration on one or more outside tasks, has no branches}
\item{Branch Taken No Dependency - one branch which will actually be taken, with no dependency for branch resolution }
\item{Branch Not Taken No Dependency - one branch which will actually be not taken, with no dependency for branch resolution}
\item{Branch Taken Dependency - one branch which will actually be taken, has a dependency for branch resolution}
\end{enumerate}
These benchmarks have been made with a purpose of analyzing performance of our proposed HTS in various basic scenarios.

 

Fig ~\ref{fig:custom1} illustrates these scenarios. Note that the plot has been normalized by the maximum value. In Fig ~\ref{fig:custom1}, one can clearly observe \textit{Naive Scheduling} performing the worst, as expected. This can be attributed to it being constrained by in-order execution and having to incur interrupt latency overhead for each task. 

\begin{algorithm}
\SetAlgoLined
\KwResult{Compressed audio}
 fetch audio\;
 Correlate audio\;
 \eIf{correlated audio $>$ threshold}{
 	\emph{time domain}\;
 	\For{$i\leftarrow 1$ \KwTo $Bands$}{
        FIR\;
        FIR\;
        FIR\;
 }
 }{
 	\For{$i\leftarrow 1$ \KwTo $Bands$}{
    	\emph{frequency domain}\;
 		FFT\;
        Vector Dot\;
        Vector Dot\;
        Vector Dot\;
        iFFT\;
      }
 }
 \caption{Audio Compression}
\end{algorithm}

There is a clear improvement in performance for both \textit{Runtime (Software) based scheduling} and \textit{HTS Scheduling} when they have multiple Functional Units(FUs), as both of them can execute tasks out-of-order. This is also true for loop based benchmarks, as iterations are also implicitly executed out-of-order (as long as they are independent).

Our speculation naively assumes branch as \textit{not taken}. So, in Fig ~\ref{fig:custom2}, first and third plots illustrate cases where \textit{HTS Scheduling} mis-speculates while in second case it speculates correctly. Notably, we observe minimal difference in performance in \textit{HTS w/o Spec} and \textit{HTS w/ Spec} when \textit{HTS Scheduling} mis-speculates (scale of the plot hides actual difference), which can be attributed to the efficient implementation of speculation. Also, \textit{HTS Scheduling} gains much better performance than \textit{Runtime (Software) based scheduling} when it speculates correctly. We assume that \textit{Runtime (Software) based scheduling} cannot speculate. 

\subsubsection*{A real life application}
We illustrate the performance of our proposed HTS on a real application. We choose Audio Compression application for the same.

Note that we are able to decompose the application into the available set of Functions. The algorithm uses FIRs (time domain) or FFTs and Vector Dot (frequency domain) based on comparison of correlated audio to a threshold.  

\begin{figure}[!t]
  \begin{centering}
  \includegraphics[width=\columnwidth]{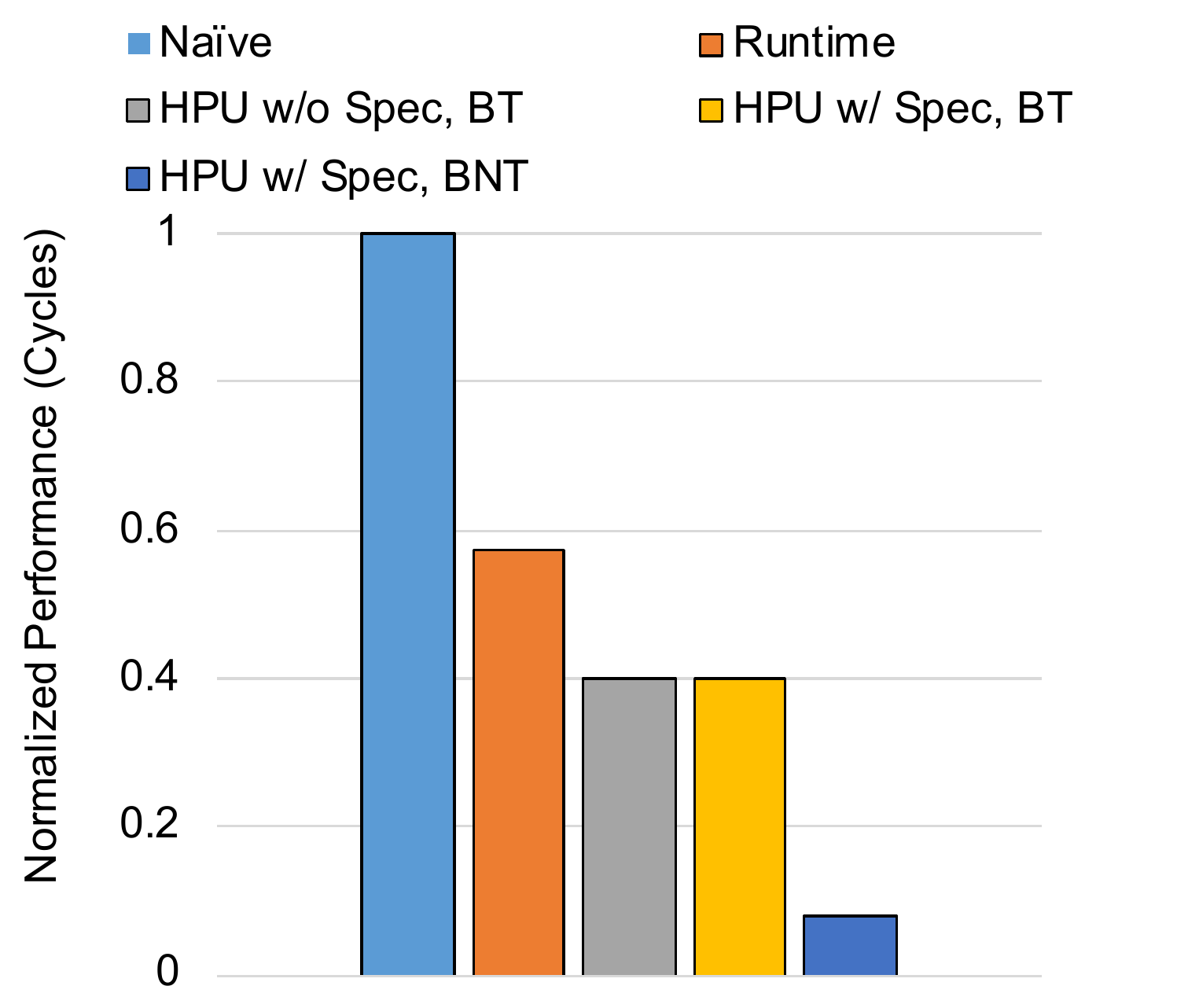}
  \caption{\label{fig:audio1}
            \footnotesize 
             Performance comparison of scheduling algorithms on audio compression.
             }
    \label{fig:audio1}
  \end{centering}
\end{figure}

Fig ~\ref{fig:audio1} depicts comparative performance of audio compression. As expected, \textit{Naive Scheduling} and \textit{Runtime (Software) based scheduling} perform poorly as compared to \textit{HTS Scheduling} owing to interrupt latency and software scheduling overheads$+$interrupt latency overheads. 

A notable difference between this application and our custom-made benchmarks is that branch resolution result drastically impacts runtime, as task blocks are different (FFT has considerably higher cycle count as compared to others). So, \textit{BT} and \textit{BNT} cases have different cycle counts.

Fig ~\ref{fig:audio2} sketches out performance trend when employing strong scaling. Hyper-parameter for this experiment is the number of iterations (Number of Bands). We change it to alter the number of tasks in the system. We observe a decrease in cycle count (increase in performance) as number of Functional Units increase in the system, since \textit{HTS Scheduling} is able to schedule tasks out-of-order. The improvement in performance is higher for program containing higher number of tasks. This is a fairly good indicator of how well our proposed HTS is performing. 

\begin{figure}[!t]
  \begin{centering}
  \includegraphics[width=\columnwidth]{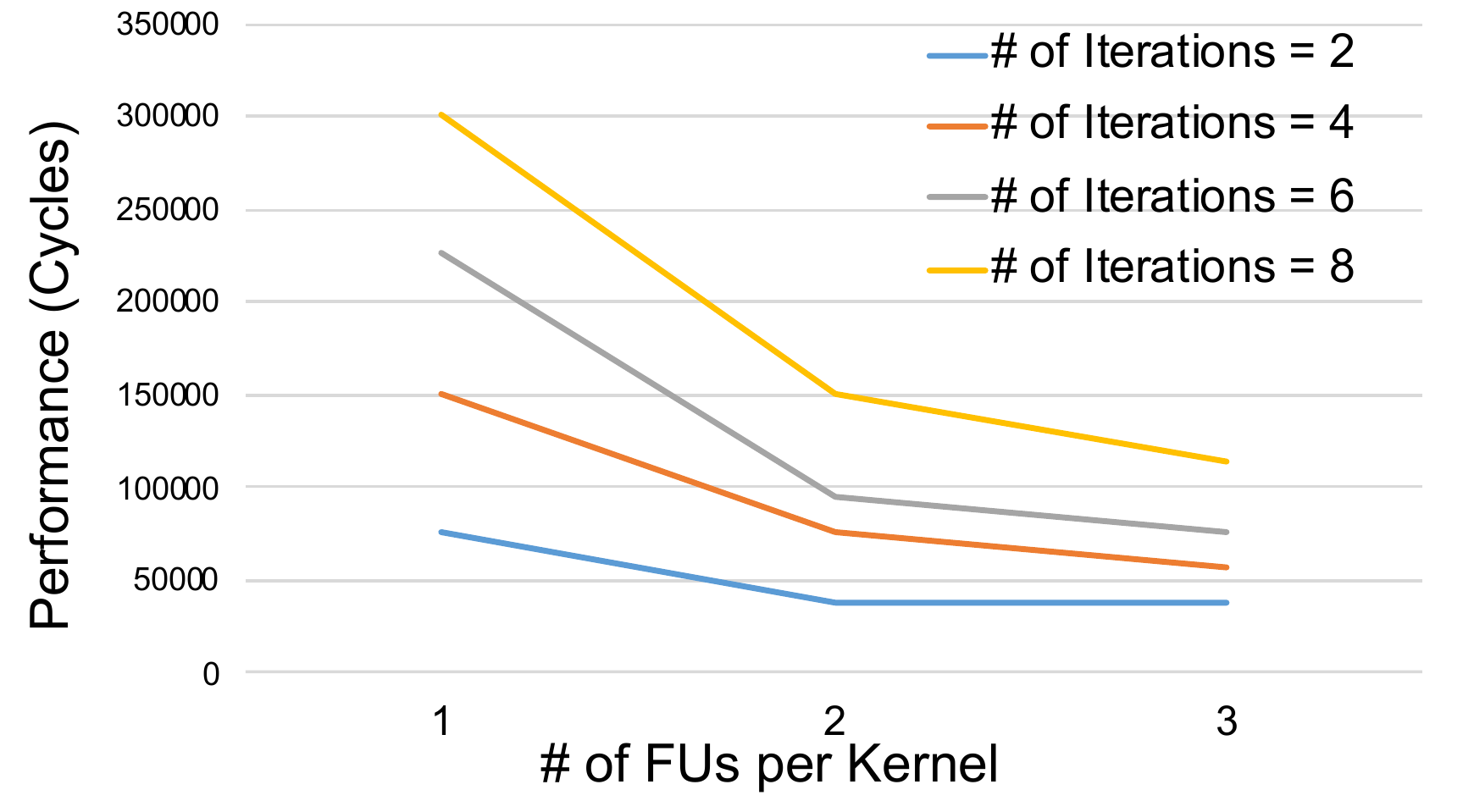}
  \caption{\label{fig:audio2}
            \footnotesize 
             Performance scaling with number of FUs on audio compression.
             }
    \label{fig:audio2}
  \end{centering}
\end{figure}

\section{Conclusion}
\label{sec:conclusion}
In this paper, we proposed to design a massively heterogeneous system architecture with a large number of accelerators. We proposed to implement accelerators at \emph{Function} abstraction, rather than \emph{Application} or \emph{Kernel} abstraction. This helps to share the accelerators across several applications, which are broken down into common set of Functions. In these scenarios, effectively scheduling the tasks in run-time becomes very crucial. We proposed to implement a hardware task scheduler in the lines of out-of-order speculative processor, that helps to overlap and execute tasks more efficiently for higher resource utilization. Our preliminary results show a great potential to provide huge uplifts in several real-life workloads.

\bibliographystyle{IEEEtran.bst}
\end{document}